\documentclass[sigconf, authorversion, nonacm]{acmart}

\acmBooktitle{}

\usepackage{graphicx} 
\usepackage{subfig}
\usepackage[most]{tcolorbox} 
\usepackage{csquotes}

\usepackage{fontawesome5}
\usepackage{tikz}
\newcommand*\circled[1]{\tikz[baseline=(char.base)]{
            \node[shape=circle,draw,inner sep=0.7pt] (char) {#1};}}

\newtcolorbox{myquote}[1][]{
    colback=black!3, 
    colframe=black!5,
    notitle,
    sharp corners,
    borderline west={2pt}{0pt}{black!80!black},
    enhanced,
    breakable,
    top=0.5pt,
    bottom=0.5pt
}

\title{GAISSALabel: A tool for energy labeling of ML models}

\author{Pau Duran, Joel Castaño, Cristina Gómez, Silverio Martínez-Fernández}
\affiliation{%
  \institution{Universitat Politècnica de Catalunya}
  \city{} 
  \country{} 
}

\email{pau.duran.manzano, joel.castano, cristina.gomez, silverio.martinez@upc.edu}

\begin{abstract}

\textit{Background}: The increasing environmental impact of Information Technologies, particularly in Machine Learning (ML), highlights the need for sustainable practices in software engineering. The escalating complexity and energy consumption of ML models need tools for assessing and improving their energy efficiency.
\textit{Goal}: This paper introduces GAISSALabel, a web-based tool designed to evaluate and label the energy efficiency of ML models. 
\textit{Method}: GAISSALabel is a technology transfer development from a former research on energy efficiency classification of ML, consisting of a holistic tool for assessing both the training and inference phases of ML models, considering various metrics such as power draw, model size efficiency, CO\textsubscript{2}e emissions and more.
\textit{Results}: GAISSALabel offers a labeling system for energy efficiency, akin to labels on consumer appliances, making it accessible to ML stakeholders of varying backgrounds. The tool's adaptability allows for customization in the proposed labeling system, ensuring its relevance in the rapidly evolving ML field.
\textit{Conclusions}: GAISSALabel represents a significant step forward in sustainable software engineering, offering a solution for balancing high-performance ML models with environmental impacts. The tool's effectiveness and market relevance will be further assessed through planned evaluations using the Technology Acceptance Model.

\end{abstract}

\makeatletter
\def\@copyrightspace{\relax}
\makeatother
\begin{document}

\maketitle

\section{Introduction}

In the rapidly evolving realm of Machine Learning (ML), a pressing challenge is the significant energy consumption of these technologies \cite{wu2022sustainable}. This trend is leading to an increased environmental footprint, as current ML development often prioritizes accuracy over energy efficiency \cite{schwartz2020green}. Addressing environmental sustainability of ML is crucial, not only for its immediate impact but also for its integral role in shaping the software landscape of the future. 

In response, building on the foundations about an energy efficiency classification for ML laid in a previous research \cite{castano2023exploring}, we present an innovative web-based tool for evaluating and labeling the energy efficiency of ML models. This tool bridges the gap identified in our study \cite{castano2023exploring} by providing a practical, user-friendly solution for the Software Engineering (SE) challenge of assessing ML models' energy consumption. While current tools often focus on the ML lifecycle aspects (e.g., \enquote{Awesome MLOps} \cite{githubGitHubVisengerawesomemlops} and \enquote{MLOps.toys} \cite{mlopsMLOpsToys}), ML quality aspects such as fairness and bias (Fairlearn \cite{fairlearnFairlearn}, Google's What-If Tool \cite{researchWhatIfTool}), or measuring the software energy consumption  \cite{bannour2021evaluating}, there is no tool to assess the energy efficiency of any ML model. In this context, GAISSALabel aims to providing a holistic assessment, covering both the training and inference phases, and suggestions to reduce the environmental footprint of ML models. This approach provides a comprehensive perspective on the environmental impact of ML models, considering computational resources, model complexity, data usage, and operational efficiency.

GAISSALabel targets to introduce a labeling system, aligned with the ML industry needs as we can see in grey literature \cite{dhar2020carbon, mitTacklingAIs, nytimesAICould} and in the emerging ISO 20226 \cite{20226} for the environmental sustainability of ML models. This feature is akin to energy labels on consumer appliances, designed to simplify the interpretation of energy efficiency metrics for ML practitioners and foster a shared understanding of energy efficiency within the ML community.

Through its integration into ML development lifecycle, GAISSALabel represents a significant technological transfer from academic research to a practical tool. It stands as a testament to the importance of sustainable practices in the field of SE, specifically in the development and deployment of high-performance ML models. GAISSALabel not only addresses the immediate need for environmental consciousness in ML but also sets a precedent for future innovations in sustainable SE, complementing initiatives like the Impact Framework by the Green Software Foundation \cite{greensoftwareWelcomeImpact}, with its specific focus on the unique challenges of ML.

\textbf{Tool code availability and screencast statement}: The code and detailed documentation is available in a replication package hosted on GitHub \faIcon{github} \cite{githubGitHubGAISSAUPCMLGAISSALabel} and Zenodo \cite{duran_manzano_2024_10532066}. The demonstration video is available on YouTube \faIcon{youtube} \cite{youtuGAISSALabelTool}. The tool is available on \cite{GAISSALabelTool}.


\section{Related work}
\label{related_work}


In the broader context of SE, initiatives from the Green Software Foundation, such as the Software Carbon Intensity specification \cite{greenSoftwareImpact} and the Impact Framework \cite{greensoftwareIF}, propose methodologies for establishing carbon emissions baselines and offer a comprehensive approach to model, measure, simulate and monitor the environmental impacts of software. EcoSoft presents an Eco-Label for Software Sustainability, aiming to integrate sustainability as a core quality of software, similar to performance or reliability \cite{deneckere2020ecosoft}. Tools like CodeCarbon \cite{CodeCarbon} and Carbontracker \cite {anthony2020carbontracker} facilitate the estimation of CO\textsubscript{2} emissions in computational tasks, providing insights into the environmental impact of software operations \cite{Cruz2022}.

Focusing on ML, studies like those by García-Martín et al. \cite{Garcia-Martin2019} and \citet{del2023dl} have explored energy consumption patterns in ML model training, offering guidelines and empirical insights into the relationship between model architectures, energy efficiency, and environmental impact. Adding to this, \citet{luccioni2023power} compares inference costs of task-specific versus general-purpose ML models, emphasizing the higher energy and carbon footprint of the latter. Other initiatives, like DENT \cite{shanbhag2023dent}, tags Stack Overflow posts with energy patterns to increase developer awareness of energy-saving opportunities in deep learning. The Green Algorithms project \cite{greenalgorithmsGreenAlgorithms} and tools provided by major cloud services like AWS, GCP, and Azure, assist in quantifying the emissions associated with ML projects \cite{microsoftChartingPath}. Projects like ML.Energy \cite{MLEnergy} try to compare different ML models and rank them ordered by some specific metric. Another significant contribution in this domain is \citet{fischer2023unified} work on ELEx, which evaluates the energy efficiency of ML models during training and inference phases. However, ELEx is limited to pre-saved models, highlighting a gap in the tools available for a broader range of ML models.


GAISSALabel addresses this gap by providing a comprehensive and adaptable tool for assessing and improving the energy efficiency of ML models. Unlike ELEx, GAISSALabel is not limited to pre-saved models; it allows users to evaluate any model, including those newly created. It is deployed online, enhancing accessibility and ease of use. The tool's integration with platforms like Hugging Face, along with its extensive database of energy efficiency labels, extends its utility. Furthermore, GAISSALabel's adaptability, demonstrated by the capability to modify metrics, computation algorithms, and label scales, distinguishes it from existing tools. 


\section{GAISSAL\lowercase{abel} in the ML lifecycle}
\label{gaissalabel_overview}
\begin{figure}[!tp]
    \centering
    \includegraphics[width=\columnwidth]{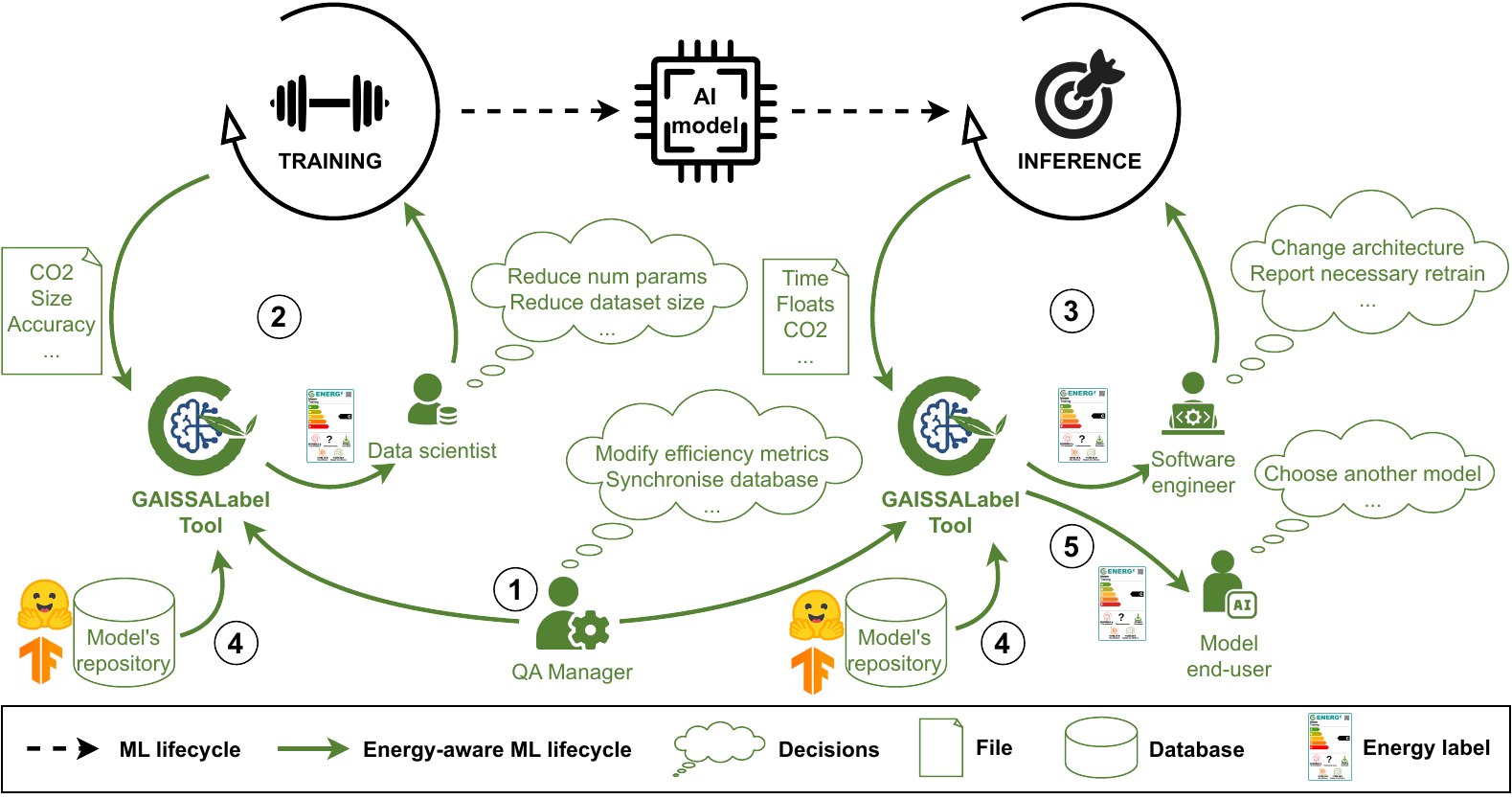}
    \caption{GAISSALabel in ML lifecycle}
    \label{fig:context_diagram}
\end{figure}

In order to address the environmental sustainability in the ML lifecycle, we propose GAISSALabel. It is a web-based tool for assisting data scientists, software engineers and end-users in measuring and improving the environmental impact of ML models, during both training and inference phases of the ML lifecycle (see Figure \ref{fig:context_diagram}). The main functionalities and challenges addressed by the proposed GAISSALabel tool are:

\textbf{Customization of the energy efficiency label definition ( \circled{1} in Figure \ref{fig:context_diagram}).} GAISSALabel provides clear and simple indication of the energy efficiency of ML models through energy efficiency labels. Currently, GAISSALabel is configured to determine the energy efficiency from a weighted average of the set of metrics defined in Table \ref{tab:metrics} and to classify this value in a A-E scale. This method of determining energy efficiency is extracted and adapted from our previous work \cite{castano2023exploring}. However, in the near future, local or international directives, or even standards may arise to propose an alternative way to compute the energy efficiency of ML models. In order to provide a tool with adaptability capabilities, GAISSALabel allows the QA manager (responsible of assuring environmental quality levels) to  customize the definition of the energy efficiency label permitting, for example, to add new metrics, change the computation method or even modify the range scale.

\textbf{Generation of energy efficiency labels of ML models (\circled{2} and \circled{3}).} GAISSALabel computes the energy efficiency of ML models and visualizes it as informative energy efficiency label, providing guidance and recommendations to potential improvements. To compute the energy efficiency, data scientists (for the training phase) or software engineers (for the inference phase) have to indicate the ML model to evaluate and to provide the values of the set of metrics (see Table \ref{tab:metrics}). Those values may be obtained from different tools, as the ones reported in Section 2 (e.g., CodeCarbon), or using the plug-in offered by GAISSALabel. Data scientists and software engineers can use the recommendations provided in energy efficiency labels to make decisions for future training (such as whether to retrain the same model or explore other models) and inference phases (such as altering the deployment architecture).

\textbf{Synchronization with any external ML model provider platform (\circled{4}).} GAISSALabel maintains a repository to save the energy efficiency labels of ML models coming from different ML model provider platforms, such as, Hugging Face and TensorFlow Hub. Currently, the repository contains 2007 models from Hugging Face. When new ML models are created in those platforms, the tool allows QA manager to trigger the creation of the energy efficiency labels for those models and save them in the repository.

\textbf{Querying energy–efficiency labels of ML models (\circled{5}).}
Allows end-users to get energy efficiency labels of the pre-saved ML models registered in the GAISSALabel repository in order to use models that make responsible energy consumption.


{\footnotesize
\begin{table}[!pb]
\centering
\begin{tabular}{|l|c|c|p{4cm}|}
\hline
\textbf{Metric} & \textbf{T} & \textbf{I} & \textbf{Description} \\ \hline
Energy consumption & X &  & Energy consumed during ML model training. \\ \hline
Downloads & X &  & Indicative of reusability; higher counts suggest efficiency. \\ \hline
Size Efficiency & X & X & Ratio of model size to CO\textsubscript{2}e. \\ \hline
Dataset Efficiency & X & X & Ratio of dataset size to CO\textsubscript{2}e. \\ \hline
Performance Score & X & X & Harmonic mean of metrics like accuracy, F1, Rouge scores. \\ \hline
CO\textsubscript{2}e & X & X & Carbon footprint during both training and inference. \\ \hline
File Size &  & X & Size of the model file. \\ \hline
Power Draw &  & X & Energy consumed per second during inference. \\ \hline
Running Time &  & X & Duration of model operation in inference. \\ \hline

FLOPS &  & X & Floating-point operations per second in inference. \\ \hline
\end{tabular}
\vspace{10pt}
\caption{Metrics used for training (T) and inference (I)}
\label{tab:metrics}
\end{table}
}

\begin{figure*}[!ht] 
    \centering
    \begin{minipage}{0.74\textwidth}
        \centering
        \includegraphics[width=0.95\textwidth]{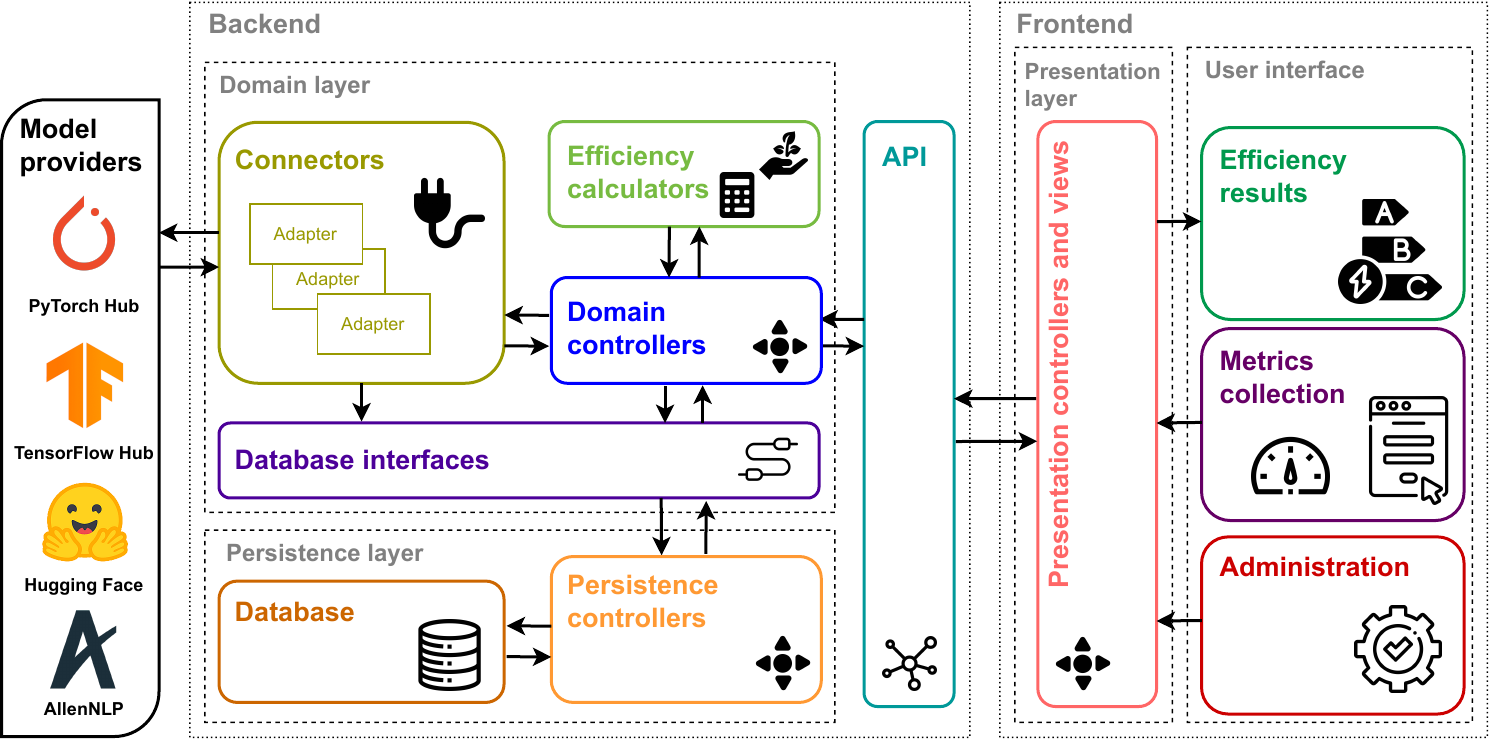} 
    \end{minipage}\hfill
    \vline\hfill
    \begin{minipage}{0.24\textwidth}
        \centering
        \includegraphics[width=0.95\textwidth]{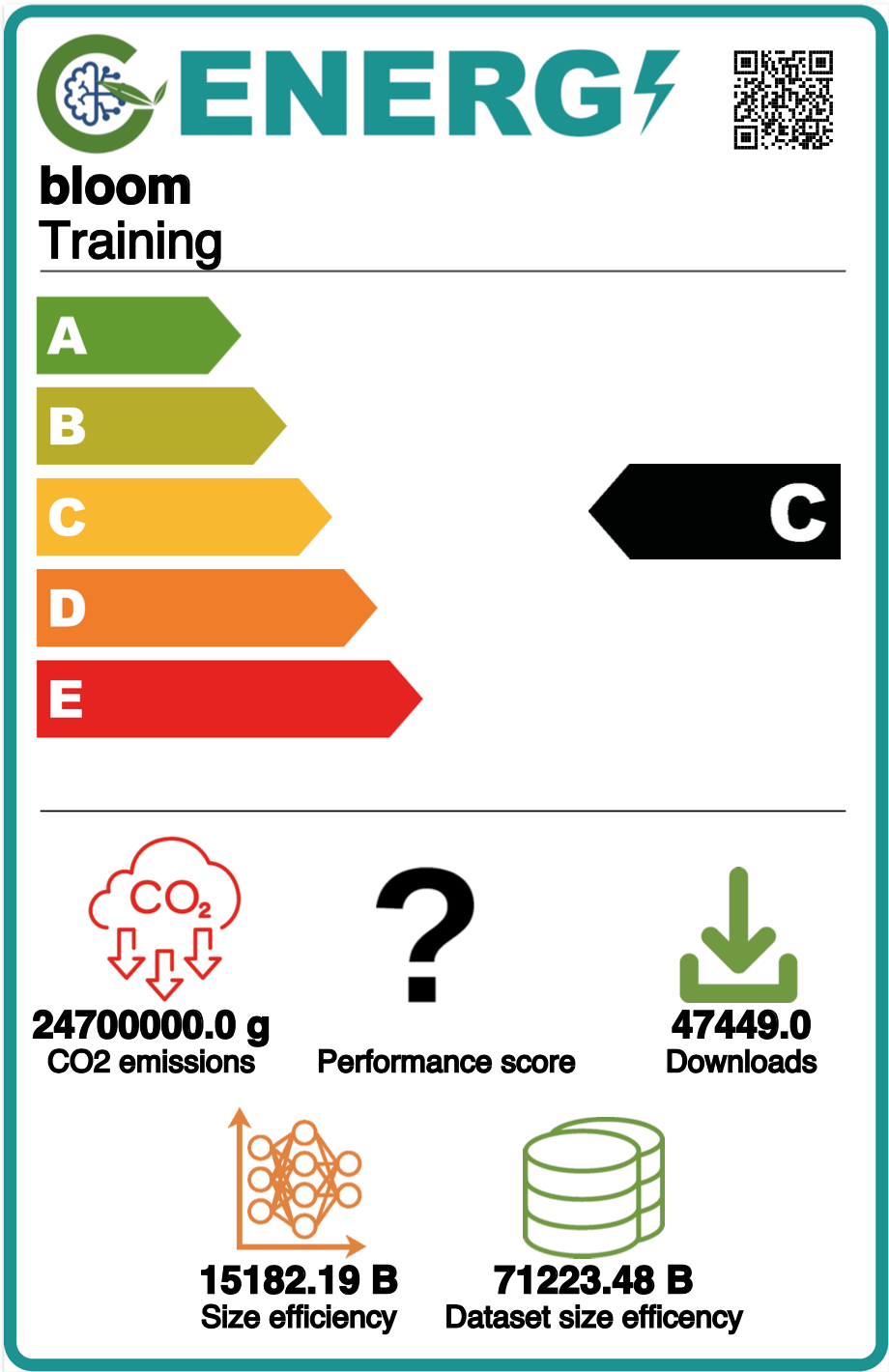} 
    \end{minipage}
    \caption{(a) The GAISSALabel Tool Architecture; (b) An example of a generated energy label for ML training}
    \label{fig:architecture}
\end{figure*}

\section{TOOL ARCHITECTURE}
\label{architecture}

 GAISSALabel has been designed as a modular, configurable and extensible tool, following a three-layer architecture (presentation, domain, and persistence layer) and a client-server architecture (frontend and backend). The resulting architecture (see Figure \ref{fig:architecture} (a)) is composed of several decoupled components capable of providing the functionalities described in Section 3 through a graphical user interface. Those components are described below.

\subsection{Frontend}

The GAISSALabel frontend comprises all the needed components for rendering the user interface.

\textbf{User interface.}
The GAISSALabel user interface has been designed to be intuitive and user-friendly. It consists of a set of web pages for presenting energy efficiency results, gathering user data, and performing QA management functions.

GAISSALabel presents energy efficiency labels (see functionalities \circled{2}\circled{3}\circled{5} in Figure \ref{fig:context_diagram}), with a comprehensible interpretation of the results for users from diverse backgrounds. As illustrated in Figure \ref{fig:architecture} (b), the label provides a conclusive assessment of energy efficiency from their metrics, along with the individual ratings for each metric. In cases where any metric is missing, the label demonstrates flexibility by omitting it and displaying a question mark \faIcon{question} instead. Furthermore, GAISSALabel shows explanations about potential implications and strategies to enhance the energy efficiency.

GAISSALabel user interface allows data scientists and software engineers to provide the data required by functionalities \circled{2} and \circled{3} in different ways. They can \textit{a)} utilize a form to indicate all the metrics values, \textit{b}) upload a file generated by external tools (e.g., CodeCarbon) with the required information or \textit{c}) use, for the inference phase, a one-click label acquisition process  where the model's deployment location and some input data to run the model are provided to present the label after conducting the inferences and collecting the results. 

Some web pages are restricted to the QA Manager (see functionalities \circled{1} and \circled{4} in Figure \ref{fig:context_diagram}) to perform the synchronization with external model providers platforms and customize the definition of the energy efficiency label.

\textbf{Presentation controllers and views.} This module comprises two primary components. Firstly, the views bear the responsibility of rendering the user interface and managing user interactions. Each web page outlined in the user interface is overseen by a dedicated view within this module. Conversely, presentation controllers are tasked with constructing these views. Additionally, they serve as coordinators bridging the specific views and the application logic delineated in the domain layer of the backend.

\subsection{Backend}
The GAISSALabel backend contains the components that run in a web server.

\textbf{API.} The API serves as the way which the frontend can request different functionalities provided by the backend. The structure of the endpoints follows the REST format and the HTTP protocol, enabling the typical GET, POST, PUT, PATCH and DELETE operations. A comprehensive list of all available endpoints can be accessed through the documentation page of the GAISSALabel server \cite{GAISSALabelAPI}. This component is also responsible for credential validation for functionalities exclusively accessible by the QA manager.

\textbf{Domain controllers.} Every API endpoint is associated with a domain controller. These domain controllers receive and process incoming requests. Through the processing, they  use some other domain components, such as when assessing the efficiency, when connecting to external model providers or when accessing to the database repository.

\textbf{Efficiency calculators.} This component assesses the energy efficiency of ML models during training and inference. Using an index-based classification system, as outlined in \cite{eu2019energylabel} and \cite{fischer2023unified}, it facilitates a balanced comparison of models by normalizing metrics against reference values. Moreover, the component classifies the energy efficiency of models into five efficiency labels from E (least efficient) to A (most efficient), based on a weighted calculation of indexed metrics. For an in-depth explanation of the efficiency calculation methodology and customization options (including weights), refer to the original study \cite{castano2023exploring}.

\textbf{Connectors.} The \textit{Connectors} component is responsible for data extraction and preprocessing, crucially retrieving data from ML model provider platforms. This component defines an adapter for each platform to collect a range of metrics for the models registered in the platform (total size of datasets used, hardware employed for training, various evaluation metrics, tags and textual descriptions associated with each model, among others). Those metrics facilitates the energy efficiency calculations of ML models. Furthermore, the adapters enable continuous updating of GAISSALabel’s internal database repository, incorporating new and refreshed data from provider platforms to expand and update the collection of ML models with energy labels. Currently, GAISALabel provides the adapter for Hugging Face but new adapters may be easily added.

\textbf{Database interfaces.} This component offers a set of interfaces to make the information persistent in the database repository. These interfaces ensure the absence of coupling between the domain and persistence layers enabling the substitution of the database management system without impacting on the rest of the architecture.

\textbf{Persistence controllers and Database}. Persistence controllers implement the database interfaces. These controllers directly interact with the database. The database design is conceived to facilitate the flexibility and adaptability of GAISSALabel.

Throughout the tool design and implementation, deliberate incorporation of environmentally conscious green software patterns (e.g., manual synchronization on demand, size reduction and log suppression \cite{calero2021software}) has been undertaken. The primary objective has been to mitigate the environmental footprint of the tool.

\section{Planned Evaluation}
\label{evaluation}

The evaluation plan has the following goal, defined applying the Goal/Question/Metric (GQM) method \cite{caldiera1994goal}:

\begin{myquote}
\textbf{Goal:} \textit{Analyze the generated energy labels and the GAISSALabel tool for the purpose of assessment with respect to ease of use and usefulness from the perspective of data scientists and software engineers in the context of energy-aware ML model training and inference.}
\end{myquote}



\textit{Sampling}: Our approach includes sampling a diverse group of software engineers and data scientists. This will ensure a representative understanding of user needs and perspectives in different organizational contexts.

\textit{Evaluation plan}: It is divided into two steps. First, engaging users in the practical application of the GAISSALabel tool for generating energy labels and assessing ML models. The usage of the tool shall be enough to go to the next step. Second, running questionnaires and semi-structured interviews. Using the Technology Acceptance Model (TAM) to gather qualitative feedback on the ease of use and usefulness \cite{davis1989perceived} of the tool. This assessment will be directly based on users' experiences with the tool during the tasks performed in the previous phase. This will involve questionnaires and interviews focused on user experience, with the aim of identifying areas for improvement in the tool.

\textit{Formative evaluation}: The outcomes of this evaluation will be instrumental in providing insights into user acceptance and practical utility of the GAISSALabel tool. These insights will drive formative evaluation, guiding future enhancements, ensuring the tool not only meets the technical requirements of ML model assessment but also aligns with the evolving market needs and user expectations.

\textit{Early feedback from previous evaluations}: 
To get early feedback on the preliminary version of  GAISSALabel tool, we conducted a questionnaire in June 2023 \cite{castano2023greenability} and a focus group in October 2023. It highlighted the need for a new version of the tool (presented in this paper), improving scalability, which motivated moving from Streamlit to 
a web-based tool, and allowing customization of the classification model capabilities. To improve understandability, we include now explanations of the evaluation and added detailed tooltips on each parameter; as well as possible directions to reduce energy consumption. Despite this early feedback, the next step will be the planned evaluation aforementioned.
 
\section{Conclusions and future work}

This paper introduces GAISSALabel, a web-based tool for energy efficiency labeling of ML models. Addressing a significant gap in the ML community's sustainable practices, GAISSALabel offers an intuitive method to assess the environmental impact of ML models. Its labeling system, which can be customized to be aligned with future standards like ISO 20226, simplifies the understanding and communication of energy efficiency for ML models.

GAISSALabel distinguishes itself with its holistic assessment approach, providing a comprehensive evaluation that covers both training and inference phases. This approach, which accounts for a range of metrics from computational requirements to CO\textsubscript{2}e emissions, offers a broader perspective on environmental sustainability in ML, setting it apart from existing tools. Furthermore, its integration into ML lifecycle exemplifies a significant technological transfer from academic research \cite{castano2023exploring} to practical application, highlighting its role in promoting sustainable SE practices.

The tool will also be engaged with the community actively through its evaluation plan, using TAM and direct engagement with software engineers and data scientists. This continuous interaction will be instrumental in refining GAISSALabel to meet evolving user needs.

Future work of GAISSALabel will focus on creating domain-specific energy labels for domains like NLP and Computer Vision, reflecting the requirements of each field. Collaboration with the ML and environmental sustainability communities will be crucial for developing standards, shared research, and tool improvements. 

In conclusion, GAISSALabel represents an advancement in integrating environmental sustainability into ML practices. Its ongoing development is expected to contribute significantly towards more energy efficient and environmentally conscious ML solutions.

\section*{ACKNOWLEDGMENTS}
The authors thank all GAISSA project members. This work is supported by TED2021-130923B-I00, funded by MCIN/AEI/10.13039/ 501100011033 and the European Union Next Generation EU/PRTR.

\bibliographystyle{IEEEtranN}

\bibliography{References}

\end{document}